\documentclass[conference,a4paper]{IEEEtran}

\usepackage{balance}
\usepackage[utf8]{inputenc} 
\usepackage[T1]{fontenc}
\usepackage{url}              % provides \url{...}
\usepackage{cite}             % improves presentation of citations

\usepackage[cmex10]{amsmath}  % Use the [cmex10] option to ensure complicance
\usepackage{amssymb,mathrsfs}                          
                              
\usepackage{mleftright}       % fix to wrong spacing of \left-,
\mleftright                   % \middle- \right-commands 

\usepackage{graphicx}         % provides \includegraphics{...} to
 \usepackage{subfig}
\usepackage{booktabs}         % fixes poor spacing in tables and
\newtheorem{lem}{Lemma}
\newtheorem{thm}{Theorem}[lem]
\newtheorem{rem}{Remark}
\newtheorem{cor}{Corollary}[thm]

\usepackage[nomain,acronym,automake,toc,nopostdot]{glossaries}
\makeglossaries
\newacronym[description=Additive white Gaussian noise]{awgn}{AWGN}{additive white Gaussian noise}
\newacronym[description= Alternating direction method of multipliers]{admm}{ADMM}{ alternating direction method of multipliers}
\newacronym[description=Approximate message passing]{amp}{AMP}{approximate message passing}
\newacronym[description={\em a posteriori} probability]{app}{APP}{{\em a posteriori} probability}
\newacronym[description= Alternative unbiased-preconditioning for iterative mMIMO
detection]{aupid}{AUPID}{alternative unbiased-preconditioning for iterative mMIMO
detection}
\newacronym[description=Base station]{bs}{BS}{base station}
\newacronym[description=Base station sleeping ]{bss}{BSS}{base station sleeping }
\newacronym[description=Belief propagation]{bp}{BP}{belief propagation}
\newacronym[description=Binary phase shift keying]{bpsk}{BPSK}{binary phase shift keying}
\newacronym[description=Bit error rate]{ber}{BER}{bit-error-rate}
\newacronym[description=Block error rate]{bler}{BLER}{block error rate}
\newacronym[description=Central limit theorem]{clt}{CLT}{central limit theorem}
\newacronym[description=Channel state information ]{csi}{CSI}{channel state information }
\newacronym[description=Closest vector problem]{cvp}{CVP}{closest vector problem}
\newacronym[description=Code division multiple access]{cdma}{CDMA}{code division multiple access}
\newacronym[description=Distributed linear data fusion]{dldf}{DLDF}{distributed linear data fusion}
\newacronym[description=European Cooperation in Science and Technology]{cost}{COST}{Cooperation in Science and Technology}
\newacronym[description=Coordinated multi-point ]{CoMP}{CoMP}{coordinated multi-point }
\newacronym[description=Correlation-based stochastic model]{cbsm}{CBSM}{correlation-based stochastic model}
\newacronym[description=Cumulative distribution function]{cdf}{CDF}{cumulative distribution function}
\newacronym[description=Degrees of freedom]{dof}{DoF}{degrees of freedom}
\newacronym[description=Element-based lattice reduction]{elr}{ELR}{element-based lattice reduction}
\newacronym[description=Extremely large aperture array]{elaa}{ELAA}{extremely large aperture array}
\newacronym[description=Fifth-generation]{5g}{5G}{fifth-generation}
\newacronym[description=Fixed-complexity sphere decoder]{fcsd}{FCSD}{fixed-complexity sphere decoder}
\newacronym[description=Forward error corrrection]{fec}{FEC}{forward error correction}
\newacronym[description=Free space path loss]{fspl}{FSPL}{free space path loss}
\newacronym[description=Gauss-Seidel]{gs}{GS}{Gauss-Seidel}
\newacronym[description=Global system for mobile communication]{gsm}{GSM}{global system for mobile communication}
\newacronym[description=Geometry-based stochastic model]{gbsm}{GBSM}{geometry-based stochastic model}
\newacronym[description=Hermite-Korkin-Zolotarev]{hkz}{HKZ}{Hermite-Korkin-Zolotarev}
\newacronym[description=Independent and identically distributed]{iid}{i.i.d.}{independent and identically distributed}
\newacronym[description=Independent and non-identically distributed]{ind}{i.n.d.}{independent and non-identically distributed}
\newacronym[description=Integer least-squares]{ils}{ILS}{integer least-squares}
\newacronym[description=Iterative discrete estimation]{ide}{IDE}{iterative discrete estimation}
\newacronym[description=Iterative discrete estimation 2]{ide2}{IDE2}{iterative discrete estimation 2}
\newacronym[description=Inter-symbol interference]{isi}{ISI}{inter-symbol interference}
\newacronym[description=Jacobi iteration]{ji}{JI}{Jacobi iteration}
\newacronym[description=International Telecommunication Union Radiocommunication Sector ]{itu-r}{ITU-R}{International Telecommunication Union Radiocommunication Sector}
\newacronym[description=Large system behaviour]{lsb}{LSB}{large system behaviour}
\newacronym[description=Lattice reduction]{lr}{LR}{lattice reduction}
\newacronym[description=Lenstra-Lenstra-Lov\'{a}sz]{lll}{LLL}{Lenstra-Lenstra-Lov\'{a}sz}
\newacronym[description=Likelihood ascent search]{las}{LAS}{likelihood ascent search}
\newacronym[description=Line-of-slight]{los}{LoS}{line-of-slight}
\newacronym[description=List sphere decoder]{lsd}{LSD}{list sphere decoder}
\newacronym[description=Linear minimum mean square error]{lmmse}{LMMSE}{linear minimum mean square error}
\newacronym[description=large multiple-input multiple-output]{lmimo}{L-MIMO}{large multiple-input multiple-output}
\newacronym[description=Log-likelihood ratio]{llr}{LLR}{log-likelihood ratio}
\newacronym[description=Long-term evolution ]{lte}{LTE}{long-term evolution}
\newacronym[description=Low density parity check]{ldpc}{LDPC}{low density parity check}
\newacronym[description=Massive machine type communications]{mmtc}{mMTC}{massive machine type communications}
\newacronym[description=Massive multiple-input multiple-output]{mmimo}{mMIMO}{massive multiple-input multiple-output}
\newacronym[description=Maximum {\em a posteriori}]{map}{MAP}{maximum {\em a posteriori}}
\newacronym[description=Maximum likelihood]{ml}{ML}{maximum likelihood}
\newacronym[description=Maximum likelihood detection]{mld}{MLD}{maximum likelihood detection}
\newacronym[description=Maximum likelihood sequence detection]{mlsd}{MLSD}{maximum likelihood sequence detection}
\newacronym[description=Maximum-ratio combining]{mrc}{MRC}{maximum-ratio combining}
\newacronym[description=Multiple-input multiple-output]{mimo}{MIMO}{multiple-input multiple-output}
\newacronym[description=Matched filter]{mf}{MF}{matched filter}
\newacronym[description=Matched-filter bound]{mfb}{MFB}{matched-filter bound}
\newacronym[description=Mean square error]{mse}{MSE}{mean square error}
\newacronym[description=Minimum mean square error]{mmse}{LMMSE}{minimum mean square error}
\newacronym[description=Mobile and wireless communications Enablers for the Twenty-twenty Information ]{metis}{METIS}{Mobile and wireless communications Enablers for the Twenty-twenty Information}
\newacronym[description=Non-line-of-sight]{nlos}{NLoS}{non-LoS}
\newacronym[description=Non-deterministic polynomial-time hard]{nphard}{NP-hard}{non-deterministic polynomial-time hard}
\newacronym[description=One dimensional]{1d}{1-D}{one dimensional}
\newacronym[description=Orthogonality defect]{od}{OD}{orthogonality defect}
\newacronym[description=Pairwise error probability]{pep}{PEP}{pairwise error probability}
\newacronym[description=Parallel interference cancellation]{pic}{PIC}{parallel interference cancellation}
\newacronym[description=Preconditioning for iterative mMIMO detection]{pid}{PID}{preconditioning for iterative mMIMO detection}
\newacronym[description=Probabilistic data association]{pda}{PDA}{probabilistic data association}
\newacronym[description=Probability distribution function]{pdf}{p.d.f.}{probability density function}
\newacronym[description=Probability mass function]{pmf}{PMF}{probability mass function}
\newacronym[description=Quadrature amplitude modulation]{qam}{QAM}{quadrature amplitude modulation}
\newacronym[description=Quadrature phase shift keying]{qpsk}{QPSK}{quadrature phase shift keying}
\newacronym[description=Receiver-side channel state information]{rcsi}{R-CSI}{receiver-side channel state information}
\newacronym[description=Received signal strength]{rss}{RSS}{received signal strength}
\newacronym[description=Regularized zero-forcing]{rzf}{RZF}{regularized zero-forcing}
\newacronym[description=Semidefinite relaxation]{sdr}{SDR}{semidefinite relaxation}
\newacronym[description=Seysen's algorithm]{sa}{SA}{Seysen's algorithm}
\newacronym[description=Signal-to-interference-plus-noise ratio]{sinr}{SINR}{signal-to-interference-plus-noise ratio}
\newacronym[description=Signal to interference ratio]{sir}{SIR}{signal to interference ratio }
\newacronym[description=Signal-to-noise ratio]{snr}{SNR}{signal-to-noise ratio}
\newacronym[description=Single antenna interference cancellation]{saic}{SAIC}{single antenna interference cancellation}
\newacronym[description=Single input single output]{siso}{SISO}{single input single output}
\newacronym[description=Singular value decomposition ]{svd}{SVD}{singular value decomposition }
\newacronym[description=Sixth-generation mobile networks]{6g}{6G}{sixth-generation}
\newacronym[description=Sphere decoder]{sd}{SD}{sphere decoder}
\newacronym[description=Space-time codes]{stc}{STC}{space-time codes}
\newacronym[description=State-of-the-art]{sota}{SoTA}{state-of-the-art}
\newacronym[description=Successive interference cancellation]{sic}{SIC}{succesive interference cancellation}
\newacronym[description=Symbol error rate]{ser}{SER}{symbol error rate}
\newacronym[description=Symmetric successive over-relaxation]{ssor}{SSOR}{symmetric successive over-relaxation}
\newacronym[description=Tabu search]{ts}{TS}{tabu search}
\newacronym[description=Three-dimensional]{3d}{3-D}{three-dimensional}
\newacronym[description=The 3rd Generation Partnership Project]{3gpp}{3GPP}{the 3rd Generation Partnership Project}
\newacronym[description=Two-dimensional]{2d}{2-D}{two-dimensional}
\newacronym[description=Uniform linear array]{ula}{ULA}{uniform linear array}
\newacronym[description=Urban micro]{umi}{UMi}{urban micro}
\newacronym[description=User equipment]{ue}{UE}{user equipment}
\newacronym[description=Vector error rate]{ver}{VER}{vector error rate}
\newacronym[description=Vertical Bell Labs layered space-time]{vblast}{V-BLAST}{vertical Bell Labs layered space-time}
\newacronym[description=Visibility region]{vr}{VR}{visibility region}
\newacronym[description=Widely linear]{wl}{WL}{widely linear}
\newacronym[description=Widely linear zero forcing]{wlzf}{WLZF}{widely linear zero forcing}
\newacronym[description=Wide-sense stationary uncorrelated scattering]{wssus}{WSSUS}{wide-sense stationary uncorrelated scattering}
\newacronym[description=Wireless World Initiative New Ratio]{winner}{WINNER}{Wireless World Initiative New Ratio}
\newacronym[description=Zero-forcing]{zf}{ZF}{zero-forcing}
\newacronym[description=Zero mean complex circularly symmetric]{zmccs}{ZMCCS}{zero mean complex circularly symmetric}

\newcommand{\figref}[1]{Fig. \ref{#1}}

\newcommand{\secref}[1]{Section \ref{#1}}

\newcommand{\thmref}[1]{{\it Theorem \ref{#1}}}
\newcommand{\corref}[1]{{\it Corollary \ref{#1}}}

\begin{document}

\title{Achieving Maximum-likelihood Detection Performance with Square-order Complexity in Large Quasi-Symmetric MIMO Systems}

%%%%%%
%\author{%
%  \IEEEauthorblockN{Anonymous Authors}
%  \IEEEauthorblockA{%
%    Please do NOT provide authors' names and affiliations\\
%    in the paper submitted for review, but keep this placeholder.\\
%    ISIT23 follows a \textbf{double-blind reviewing policy}.}
%}
 \author{
   \IEEEauthorblockN{Jiuyu Liu, Yi Ma, and Rahim Tafazolli}
   \IEEEauthorblockA{5GIC and 6GIC, Institute for Communication Systems, University of Surrey\\
                     GU2 7XH, Guildford, UK\\
                     \{jiuyu.liu, y.ma, r.tafazolli\}@surrey.ac.uk}
}       
\maketitle

\begin{abstract}
We focus on the signal detection for large quasi-symmetric (LQS) multiple-input multiple-output (MIMO) systems, where the numbers of both service ($M$) and user ($N$) antennas are large and $N/M \rightarrow 1$.
It is challenging to achieve maximum-likelihood detection (MLD) performance with square-order complexity due to the ill-conditioned channel matrix.
In the emerging MIMO paradigm termed with an extremely large aperture array, the channel matrix can be more ill-conditioned due to spatial non-stationarity.
In this paper, projected-Jacobi (PJ) is proposed for signal detection in (non-) stationary LQS-MIMO systems.
It is theoretically and empirically demonstrated that PJ can achieve MLD performance, even when $N/M = 1$.
Moreover, PJ has square-order complexity of $N$ and supports parallel computation.
The main idea of PJ is to add a projection step and to set a (quasi-) orthogonal initialization for the classical Jacobi iteration.
Moreover, the symbol error rate (SER) of PJ is mathematically derived and it is tight to the simulation results.
\end{abstract}

\section{Introduction}\label{sec01}
We focus on the signal detection for large quasi-symmetric (LQS) \gls{mimo} systems.
Its signal model can be represented in the following standard vector/matrix form
\begin{equation}\label{eqn01}
	\mathbf{y} = \mathbf{H} \mathbf{x} + \mathbf{v},
\end{equation} 
where $\mathbf{y} \in \mathbb{C}^{M \times 1}$ denotes the observed vector, $\mathbf{H} \in \mathbb{C}^{M \times N}$ ($M \geq N$) the random channel matrix, $\mathbf{x} \in \mathbb{C}^{N \times 1}$ the transmitted vector, $\mathbf{v} \sim \mathcal{CN}(0,\sigma_v^2\mathbf{I}_M)$ the \gls{awgn}, and $\mathbf{I}_M$ stands for an $(M) \times (M)$ identity matrix. 
Each element in $\mathbf{x}$ is assumed to be drawn from a finite-alphabet set $\mathcal{X}$ of size $\mathcal{J}$ with equal probability, and fulfills: $\mathbb{E}(\mathbf{x}) = \mathbf{0}$, $\mathbb{E}\left(\mathbf{x}\mathbf{x}^H\right) = \sigma_{x}^{2}\mathbf{I}$.

In LQS-MIMO systems, $M$ is assumed to be very large, e.g., hundreds or even thousands, and $N$ can be comparable to $M$, i.e., $N/M \rightarrow 1$.
In addition, the wireless channel can be spatially non-stationary due to the use of  extremely large aperture array (ELAA) \cite{Wei2022}.
All these features render the LQS-MIMO channel to be ill-conditioned.
The optimal MIMO detector, called \gls{mld}, is known for solving an NP-hard problem.
It is challenging to achieve the \gls{mld} performance with square-order complexity in an ill-conditioned channel matrix \cite{BJORNSON20193}.
Given perfect channel knowledge, current MIMO detectors can be divided into two main types: non-linear and linear.

The optimal \gls{mld} is a non-linear detector given by \cite{Damen2003}
\begin{equation}\label{eqn02}
	\widehat{\mathbf{x}}_{\textsc{mld}}=\underset{\mathbf{x}\in\mathcal{X}^{N}}{\arg \min}\ \|\mathbf{y} - \mathbf{H} \mathbf{x}\|^{2},
\end{equation}
which aims to find the closest lattice point to $\mathbf{x}$.
However, its searching space ($\mathcal{J}^{N}$) is prohibitive for large $N$.
For the sake of low-complexity, several search-based methods have been proposed with constrained searching spaces \cite{Sun2009,Wong2002,Damen2003,Kisialiou2009}.
However, they still require an exponentially grows complexity to offer (near-) MLD performance \cite{Kisialiou2009}.
Recently, \gls{admm} has been used to solve \eqref{eqn02} with constraints \cite{Souto2016,Shahabuddin2021,Tiba2021,Zhang2022}.
The serial-complexity of ADMM-based methods is $\mathcal{O}(N^3)$, mainly caused by matrix inverse.
In LQS-MIMO systems,  they have a considerable performance gap compared to MLD, due to the channel ill-conditioning.

Linear MIMO detectors mainly include \gls{mf} and \gls{rzf}.
Denote $\mathbf{b} = \mathbf{H}^{H}\mathbf{y}$ as the MF estimation,  the solution of \gls{rzf} is given by \cite{Yang2015}
\begin{equation} \label{eqn03}
	\widehat{\mathbf{x}}_{\textsc{rzf}} = \mathbf{A}^{-1} \mathbf{b},
\end{equation}
where $\mathbf{A} = \mathbf{H}^{H}\mathbf{H} + \rho \mathbf{I}$ is a Gram matrix, and $\rho \geq 0$ is the regularization parameter.
However, the serial-complexity of computing $\mathbf{A}^{-1}$ is $\mathcal{O}(N^3)$, which is prohibitive for large $N$.
A number of iterative methods have been proposed to bypass the matrix inverse, mainly including the following four types \cite{Donoho2009,Yang2015,Albreem2019, Li2022}: {\it 1)} stationary iterative methods, {\it 2)} gradient descent methods, {\it 3)} quasi-Newton methods, and {\it 4)} \gls{amp}\footnote{AMP fails in ill-conditioned channel matrix. 
To overcome this problem, several variants of AMP  have been proposed with decorrelated linear estimator \cite{Ma2017,Rangan2019,Liu2021a,Takeuchi2021,Liu2022a,Liu2022}.
However, a considerable performance gap exists between these methodologies and MLD, especially in highly loaded large-MIMO systems.}.
These iterative methods can offer the RZF detection performance with square-order complexity.
However, the detection performance of \gls{rzf} (and these iterative methods) is too sub-optimal in LQS-MIMO systems, due to the ill-conditioned channel matrix.

In this paper, projected Jacobi (PJ) is proposed for signal detection in (non-) stationary LQS-MIMO systems.
The complexity of PJ is $\mathcal{O}(N^2)$ and supports parallel computation.
The main idea of PJ is to add a projection step into the classical Jacobi iteration.
The initialization of PJ can be selected to be the decision of MF or RZF, where the latter can be obtained by low-complexity iterative methods.

It is proved that, when $M \rightarrow \infty$, MLD has asymptotically the same \gls{pep} as \gls{mfb} \footnote{\gls{mfb} is an ideal bound, where the interference is assumed to be perfectly removed and then maximum-ratio combining is performed at each interference-free channel. Some related discussions can be found in \cite{Ling1995,Baas2001,Wu2005}.} for any error pattern.
This means that the detection performance of MLD can be the same as that of MFB, e.g., \gls{ser}.
Moreover, it is theoretically and empirically demonstrated that, with an appropriate initialization, PJ can offer MFB/MLD detection performance in LQS-MIMO systems, even when $N/M = 1$.
Furthermore, the \glspl{ser} of MFB and PJ are mathematically derived and they well fit our simulation results.

\section{Channel Models, Preliminaries and Problem Statement} \label{sec02}
\subsection{Channel Models for LQS-MIMO Systems} \label{sec02a}
In LQS-MIMO systems, the elements of $\mathbf{H}$ can obey \gls{iid} Rayleigh fading as follows \cite{Hochwald2004}
\begin{equation} \label{eqn04}
h_{m,n} \sim \mathcal{CN} \Big(0, \dfrac{\sigma^{2}_{h}}{M}\Big), \ \forall m, n,
\end{equation}
where $\frac{\sigma^{2}_{h}}{M}$ denotes the variance of each channel element.
In LQS-MIMO systems, the aperture of service-antenna array can be very large and spherical-wave model should be used to describe the wireless channel, e.g., \cite{Amiri2018, Rodrigues2020,Amiri2022,Wang2023,DeCarvalho2020}.
Therefore, the elements of $\mathbf{H}$ obey i.n.d. (n. for non-identically) Rayleigh fading as follows
\begin{equation} \label{eqn05}
h_{m,n} \sim \mathcal{CN} \Big(0, w_{m,n}\dfrac{\sigma^{2}_{h}}{M}\Big) ,
\end{equation}
where $w_{m,n}$ represents the large-scale fading.
According to the measurement results (e.g, \cite{DeCarvalho2020,Harris2016,6410305,7062910,7914746}), spherical-wave model is more accurate in practical MIMO systems with large $M$.

\subsection{Jacobi Iteration} \label{sec2b}
Jacobi iteration is one of the stationary iterative methods.
It aims to find $\widehat{\mathbf{x}}_{\textsc{rzf}}$ bypassing the inversion of $\mathbf{A}$ as follows \cite{Albreem2019}
\begin{equation} \label{eqn06}
\mathbf{x}_{t+1} = \mathbf{x}_{t} + \mathbf{D}^{-1}(\mathbf{b} - \mathbf{A}\mathbf{x}_{t}),
\end{equation}
where $t \geq 0$ denotes the iteration index, and $\mathbf{D} \in \mathbb{C}^{N\times N}$ is a diagonal matrix containing the diagonal components of $\mathbf{A}$.
If the spectral radius of $(\mathbf{I} - \mathbf{D}^{-1}\mathbf{A})$ is less than $1$, $\mathbf{x}_{t}$ can surely converge to $\widehat{\mathbf{x}}_{\textsc{rzf}}$ \cite{Wang2022a}.

\subsection{PEPs of MLD}
Suppose that $\mathbf{x}$ is erroneously detected to $\mathbf{z}$, the PEP of MLD conditioned on $\mathbf{H}$ is given by \cite{Handte2009}
\begin{equation} \label{eqn07}
\mathscr{P}_{\textsc{mld}}(\mathbf{x} \rightarrow \mathbf{z} | \mathbf{H}) = \mathcal{Q}\left(\sqrt{\frac{\|\mathbf{H}\mathbf{e}\|^{2}}{2\sigma_{v}^{2}}}\right),
\end{equation}
where $\mathscr{P}_{\textsc{mld}}$ denotes the PEP of MLD, $\mathbf{e} \triangleq \mathbf{x} -\mathbf{z}$ the detection error, and $\mathcal{Q}(\cdot)$ is the Gaussian Q-function.
Suppose that there is only a single error made at the $n^{th}$ data stream to its nearest neighbor, the conditional PEP of MLD is given by \cite{Handte2009}
\begin{equation} \label{eqn08}
	\mathscr{P}_{\textsc{mld}}(\mathbf{x} \rightarrow \mathbf{z} | \mathbf{H}, n) = \mathcal{Q}\left(\sqrt{\frac{\|\mathbf{h}_{n}\|^{2} d^{2}_{\mathrm{min}}}{2\sigma_{v}^{2}}}\right),
\end{equation}
where $d_{\mathrm{min}} > 0$ represents the minimum Euclidean distance between two different symbols in the constellation set $\mathcal{X}$, and $\mathbf{h}_{n} \in \mathbb{C}^{M \times 1}$ denotes the $n^{th}$ column vector of $\mathbf{H}$.
Taking $\mathcal{J}$-QAM as an example, $d_{\mathrm{min}}$ is given by \cite{Barry2012}
\begin{equation} \label{eqn09}
	d_{\mathrm{min}} = \sqrt{\dfrac{6 \sigma_{x}^{2}}{\mathcal{J}-1}}.
\end{equation}

\subsection{Statement of the Research Problem} \label{sec02d}
According to the discussions in \secref{sec01} and \ref{sec02a}, practical LQS-MIMO detector should satisfy the following three technical requirements: {\it 1)} low complexity, e.g., $\mathcal{O}(N^2)$ or even lower, {\it 2)} (Close-to-) MLD performance, and {\it 3)} parallel computation support.
So far, there is no MIMO detector that can simultaneously fulfill the three requirements.
The following sections are therefore motivated.

\section{The Development of PJ} \label{sec03}
In this section, PJ is proposed for LQS-MIMO detection.
It can simultaneously fulfill the three technical requirements discussed in \secref{sec02d}.

\subsection{Comparison of MFB and Jacobi Iteration} \label{sec03a}
The following MFB is often used as the performance lower bound of MIMO detectors \cite{Wu2005}
\begin{equation}\label{eqn10}
\widehat{\mathbf{x}}_{\textsc{mfb}} = \mathbf{x} + \mathbf{D}^{-1}\mathbf{H}^{H}\mathbf{v},
\end{equation}
where the \gls{isi} is assumed to be perfectly removed.
Jacobi iteration can be reformulated as follows
\begin{equation}\label{eqn11}
 	\mathbf{x}_{t+1} = \mathbf{F}\mathbf{e}_{t} +  \mathbf{x} + \mathbf{D}^{-1}\mathbf{H}^{H}\mathbf{v},
\end{equation}
where $\mathbf{e}_{t} \triangleq \mathbf{x}_{t} - \mathbf{x}$ denotes the estimation error at the $t^{th}$ iteration, and $\mathbf{F} \triangleq \mathbf{I} - \mathbf{D}^{-1}\mathbf{A}$ is a hollow matrix.

Compare  \eqref{eqn10} and \eqref{eqn11}, it can be found that the only difference between them is the error-related term $\mathbf{F}\mathbf{e}_{t}$.
This means that we will have $\mathbf{x}_{t+1} = \widehat{\mathbf{x}}_{\textsc{mfb}}$ when $\mathbf{e}_{t} = \mathbf{0}$. 
Since the detection performances of MLD and MFB are asymptotically the same when $M \rightarrow \infty$ (see the proof in \secref{sec04a}).
The following section is therefore motivated.

\subsection{PJ with (Quasi-) Orthogonal Initialization} \label{sec03b}
It is hard to have $\mathbf{x}_{t} = \mathbf{x}$ if elements of $\mathbf{x}_{t}$ are continuous, since $\mathbf{x}\in \mathcal{X}^{N}$. 
Therefore, a projection step is proposed to be added in Jacobi iteration (so-called PJ) as follows
\begin{subequations} \label{eqn12}
	\begin{align}
		\mathbf{x}_{t+1} &= \mathbf{z}_{t} + \mathbf{D}^{-1}(\mathbf{b} - \mathbf{A}\mathbf{z}_{t}), \label{eqn12a}\\
		\mathbf{z}_{t+1} &= \mathcal{S}(\mathbf{x}_{t+1}) \label{eqn12b},
	\end{align}
\end{subequations}
where $\mathbf{z}_{t}$ denotes the decision in the $t^{th}$ iteration, and $\mathcal{S}(\cdot)$ is a slicing function making symbol-by-symbol decision.

For iterative methods, the initialization is usually set to be $\mathbf{0}$, i.e., $\mathbf{z}_{0} = \mathbf{0}$, but this is not always an appropriate setting for PJ in LQS-MIMO systems.
According to {\it Kashin–Tzafriri Theorem}, the column subset of $\mathbf{F}$ has smaller condition number than $\mathbf{F}$ \cite{Tropp2009}.
In \eqref{eqn11}, if $\mathbf{e}_{0}$ is a sparse vector, it is equivalent to choosing a column subset of $\mathbf{F}$. 
The problem of channel ill-conditioning can be overcome if $\mathbf{e}_{0}$ is sufficiently sparse.
Therefore, an appropriate initialization is required for PJ to guarantee the convergence.
For instance, if $\mathbf{z}_{0} = \mathbf{0}$, $\mathbf{z}_{1}$ will be the decision of MF.
In this case, the sparsity of $\mathbf{e}_{0}$ can only be guaranteed if $N \ll M$.
On the contrary, RZF can offer better detection performance in \gls{isi}-dominated MIMO channels, compared to MF \cite{fowc}. 
Therefore, in LQS-MIMO systems, the initialization of PJ can also be set as follows
\begin{equation} \label{eqn13}
	\mathbf{z}_{0} = \mathcal{S}\left(\widehat{\mathbf{x}}_{\textsc{rzf}}\right),
\end{equation}
which is called orthogonal initialization when $\rho = 0$, since the \gls{isi} is totally removed.
Otherwise $\mathbf{z}_{0}$ is called quasi-orthogonal initialization.
For instance, if $\rho = \sigma_{v}^{2}/\sigma_{x}^{2}$, \eqref{eqn03} is called \gls{mmse}, which is the optimal linear MIMO detector.
Moreover, the decision of PJ is selected to be $\mathbf{z}_{t^{\star}}$ with $t^{\star}$ as follows 
\begin{equation} \label{eqn14a}
	t^{\star} =\underset{t \in 1,..., T}{\arg \min}\ \|\mathbf{r}_{t}\|^{2},
\end{equation}
where $\mathbf{r}_{t} \triangleq \mathbf{b} - \mathbf{A}\mathbf{z}_{t}$ denotes the residual.
The objective of \eqref{eqn14a} is to output the closest lattice to $\mathbf{x}$.

\subsection{Complexity Analysis of PJ}
In \eqref{eqn12a}, the complexity of calculating $\mathbf{A}\mathbf{z}_{t}$ is $N^{2}$.
Since $\mathbf{D}$ is a diagonal matrix, the complexity of calculating $\mathbf{D}^{-1}\mathbf{r}_{t}$ is $N$.
The complexity of addition, subtraction and projection is negligible \cite{Zhang2022}.
Therefore, the complexity of PJ is $N^2 + N$, which is in the square-order of $N$.
As discussed in \secref{sec01}, some iterative methods can find $\widehat{\mathbf{x}}_{\textsc{rzf}}$ with parallel-complexity of $\mathcal{O}(N^2)$.
Therefore, the overall complexity of PJ is still in the square-order of $N$ and supports parallel computation.

\section{Performance Analysis} \label{sec4}
In this section, we first prove that MFB and MLD have the same detection performance in LQS-MIMO systems. 
Then, it is proved that PJ can achieve MFB (or MLD) performance with an appropriate initialization.
Furthermore, the SERs of MFB and PJ are mathematically derived, respectively.

\subsection{PEP and SER Analysis of MFB} \label{sec04a}
\begin{lem} \label{lem1}
Suppose that $\mathbf{x}$ is erroneously detected to $\mathbf{z}$, the PEP of MFB conditioned on $\mathbf{H}$ is as follows
\begin{equation} \label{eqn14}
\mathscr{P}_{\textsc{mfb}}(\mathbf{x} \rightarrow \mathbf{z} | \mathbf{H}) = \mathcal{Q}\left(\sqrt{\frac{\|\mathbf{e}\|^{4}}{2\sigma_{v}^{2}\|\mathbf{H}\mathbf{D}^{-1}\mathbf{e}\|^{2} }}\right),
\end{equation}
where $\mathscr{P}_{\textsc{mfb}}(\cdot)$ denotes the PEP of MFB.
\end{lem}

The following two theorems aim to show the coincidence of MFB performance and MLD performance in both stationary and non-stationary LQS-MIMO channels.

\begin{thm} \label{thm1.1}
Suppose that: {\it 1)} elements of $\mathbf{H}$ follow \gls{iid} Rayleigh distribution as \eqref{eqn04}, and {\it 2)} $\mathbf{x}$ is erroneously detected to $\mathbf{z}$, the PEPs of MFB and MLD will be asymptotically the same when $M \rightarrow \infty$ as follows
\begin{equation} \label{eqn15}
	 \left.\begin{array}{l}
		 \lim\limits_{M \rightarrow \infty}\mathscr{P}_{\textsc{mfb}}(\mathbf{x} \rightarrow \mathbf{z} | \mathbf{H}) \\
		 \lim\limits_{M \rightarrow \infty}\mathscr{P}_{\textsc{mld}}(\mathbf{x} \rightarrow \mathbf{z} | \mathbf{H})
	\end{array}\right\}
	=\mathcal{Q}\left(\sqrt{\frac{\sigma_{h}^{2}\|\mathbf{e}\|^{2}}{2\sigma_{v}^{2}} }\right).
\end{equation}
\end{thm}

\begin{thm} \label{thm1.2}
	Suppose that: {A1)} elements of $\mathbf{H}$ follow i.n.d. Rayleigh distribution as \eqref{eqn05}, {A2)} $\mathbf{x}$ is erroneously detected to $\mathbf{z}$, {A3)} the service-antenna array is equally divided into $S$ subarrays each with $M/S$ antennas, {A4)} given $n$, $w_{m,n}$ is the same for all $m$ belonging to the same subarray, and {A5)} the large-scale fading $w_{m,n}$ is normalized as follows
\begin{equation} \label{eqn16}
	\sum_{m = 1}^{M} w_{m,n} = M, \ \forall n,  
\end{equation}
the PEPs of MFB and MLD will be asymptotically the same when $M \rightarrow \infty$ as follows
	\begin{equation} \label{eqn17}
	\left.\begin{array}{l}
	\lim\limits_{M \rightarrow \infty}\mathscr{P}_{\textsc{mfb}}(\mathbf{x} \rightarrow \mathbf{z} | \mathbf{H}) \\
	\lim\limits_{M \rightarrow \infty}\mathscr{P}_{\textsc{mld}}(\mathbf{x} \rightarrow \mathbf{z} | \mathbf{H})
	\end{array}\right\}
	=\mathcal{Q}\left(\sqrt{\frac{\sigma_{h}^{2}\|\mathbf{e}\|^{2}}{2\sigma_{v}^{2} }}\right).
	\end{equation}
\end{thm}

The condition {A5} means that the average power of every user-antenna is normalized to be the same for a fair comparison.
Please note that the MIMO channel is still spatially non-stationary, since $w_{m,n}$ varies for different $m$.

\begin{rem}	
\thmref{thm1.1} and \ref{thm1.2} are hold for any error pattern when $N \leq M$.
This indicates that, MLD can offer the same the detection performance as MFB from very asymmetric MIMO (i.e., $N \ll M$ ) to symmetric MIMO (i.e., $N/M = 1$).
Compare \thmref{thm1.1} and \ref{thm1.2}, it can be found that the values of PEP are the same in stationary and non-stationary large MIMO channels.
This means that the network-side spatial non-stationarity will not degrade the detection performance of MLD.
Moreover, the PEPs are the same for any channel realization due to the channel hardening effects.
As the result, the average-SERs (in average of channel realization) of MLD and MFB  are also the same.
\end{rem}

\begin{rem}
When $M \rightarrow \infty$, each data stream of MFB can offer the same detection performance as the AWGN channel, since the \gls{isi} is perfectly removed.
According to \thmref{thm1.1} and \ref{thm1.2}, MLD can also offer the AWGN performance in LQS-MIMO systems.
This means that the performance of MLD will not be degraded by ISI.
This is counter-intuitive, but our simulation results confirm this finding (see {\it Experiment 1} in \secref{sec5}).
In LQS-MIMO systems, MLD can serve a large number of user-antennas each with AWGN performance, but suffers from prohibitive complexity.
\end{rem}

\begin{thm} \label{thm1.3}
Suppose that MFB makes only a single error at the $n^{th}$ data stream to its nearest neighbor, the conditional PEPs of MFB and MLD are the same as follows
\begin{equation} \label{eqn18}
\mathscr{P}_{\textsc{mfb}}(\mathbf{x} \rightarrow \mathbf{z} | \mathbf{H}, n) = \mathscr{P}_{\textsc{mld}}(\mathbf{x} \rightarrow \mathbf{z} | \mathbf{H}, n).
\end{equation}
\end{thm}

\thmref{thm1.3} holds for any MIMO size and channel distribution.
The nearest error usually happens at high \glspl{snr}, which is precisely the working area of LQS-MIMO systems.
\gls{ser} is commonly used to measure the reliability of a MIMO detector.
Therefore, we consider to derive a theoretical SER for MFB in MIMO systems.

\begin{cor} \label{cor1.3.1}
Denote $\mathcal{E}_{n}^{(\textsc{mfb})}$ as the error probability of user $n$, suppose that the error happens independently for all $n$, the SER of MFB, denoted by $\mathcal{E}_{\textsc{mfb}}$, is as follows
\begin{equation} \label{eqn19}
	\mathcal{E}_{\textsc{mfb}} = \dfrac{1}{N}\sum_{n=1}^{N} \mathcal{E}_{n}^{(\textsc{mfb})}.
\end{equation}
Since the explicit expression of SER is mathematical intractable \cite{fowc}, 
it is proposed to approximate $\mathcal{E}_{n}^{(\textsc{mfb})}$ based on \thmref{thm1.3}. 
Suppose that each element of $\mathbf{x}$ is drawn from $\mathcal{X}$ with equal probability, $\mathcal{E}_{n}^{(\textsc{mfb})}$ can be expressed as follows
\begin{equation} \label{eqn20}
	\mathcal{E}_{n}^{(\textsc{mfb})} \approx \mathcal{K} \mathscr{P}_{\textsc{mfb}}(\mathbf{x} \rightarrow \mathbf{z} | \mathbf{H}, n),
\end{equation}
where $\mathcal{K}$ represents the average number of nearest neighbors for the symbols in $\mathcal{X}$.
\end{cor}

For $\mathcal{J}$-QAM modulation, suppose that $\mathcal{J}$ is the integer power of $4$, $\mathcal{K}$  is given by \cite{Barry2012}
\begin{equation} \label{eqn21}
	\mathcal{K} = 4 - \dfrac{4}{\sqrt{\mathcal{J}}}.
\end{equation}

\subsection{PEP and SER Analysis of PJ} \label{sec04b}
To simplify the mathematical analysis, the iteration of PJ is ignored in this section.
Denoted $\widehat{\mathbf{x}}_{\textsc{pj}}$ as the estimation of PJ, eqn. \eqref{eqn11} can be rewritten as follows
\begin{equation} \label{eqn22}
	\widehat{\mathbf{x}}_{\textsc{pj}} = \mathbf{F} \bar{\mathbf{e}}+  \mathbf{x} + \mathbf{D}^{-1}\mathbf{H}^{H}\mathbf{v},
\end{equation}
where $\bar{\mathbf{e}} \triangleq \bar{\mathbf{z}} - \mathbf{x}$ denotes initialization error and $\bar{\mathbf{z}}$ denotes the initialization vector.

\begin{lem} \label{lem2}
	Given $\bar{\mathbf{z}}$, suppose that $\mathbf{x}$ is erroneously detected to $\mathbf{z}$, the PEP of PJ conditioned on $\mathbf{H}$ and $\bar{\mathbf{z}}$ is as follows
	\begin{equation}\label{eqn23}
		\mathscr{P}_{\textsc{pj}} \left(\mathbf{x} \rightarrow \mathbf{z} | \mathbf{H}, \bar{\mathbf{z}}\right) = \mathcal{Q}\left(\frac{\|\mathbf{e}\|^{2} + 2\Re\{\mathbf{e}^{H}\mathbf{F}\bar{\mathbf{e}}\} } {\sqrt{2\sigma^{2}_{v}  \| \mathbf{H} \mathbf{D}^{-1}\mathbf{e}\|^{2} }} \right).
	\end{equation}
\end{lem}

If there is no error at the initialization vector, i.e., $\bar{\mathbf{e}} = \mathbf{0}$, the term $2\Re\{\mathbf{e}^{H}\mathbf{F}\bar{\mathbf{e}}\}$ in the numerator is $0$, and the PEP of PJ is the same as that of MFB.
This is corresponding to the discussion in \secref{sec03a}.

\begin{thm}\label{thm3.1}
	Suppose that: {B1)} there is a single error at the $k^{th}$ data stream of $\bar{\mathbf{z}}$ to its nearest neighbor, {B2)} PJ makes only a single error at the $n^{th}$ data stream to its nearest neighbor, and {B3)} the modulation is $\mathcal{J}$-QAM, the conditional PEP of PJ is as follows
	\begin{equation} \label{eqn24}
	 \mathscr{P}_{\textsc{pj}}\left(\mathbf{x} \rightarrow \mathbf{z} | \mathbf{H}, n, k \right) = \dfrac{1}{4} \sum_{\ell = 1}^{4} \mathcal{Q}\left( \sqrt{\dfrac{ \|\mathbf{h}_{n}\|^{2}  d^{2}_{\mathrm{min}} \Gamma_{\ell, n, k}^{2} }{2 \sigma^2_{v}}} \right),
	\end{equation}	
	where 
	\begin{equation} \label{eqn25}
	\Gamma_{\ell, n, k} \triangleq 1 + 2 \Re\{\mathbf{u}_\ell \mathbf{F}_{n,k}\},
	\end{equation}
	where $\mathbf{u} \triangleq [1, -1, i, -i]^{T}$ and $i \triangleq \sqrt{-1}$ is the imaginary unit.
\end{thm}

The only different between the PEPs of PJ and MLD is the term $\Gamma_{\ell, n, k}$.
If we have $\Gamma_{\ell, n, k} = 1, \forall \ell, n, k$, the PEPs of PJ and MLD will be the same for any error pattern.

\begin{cor} \label{cor2.1.1}
Given the conditions {B1}, {B2} and {B3}, when $M \rightarrow \infty$, the conditional PEP of PJ  will be asymptotically the same as that of MFB/MLD as follows
\begin{equation}
	\lim\limits_{M \rightarrow \infty} \mathscr{P}_{\textsc{pj}}\left(\mathbf{x} \rightarrow \mathbf{z} | \mathbf{H}, n, k \right) = \mathscr{P}_{\textsc{mld}/\textsc{mfb}}(\mathbf{x} \rightarrow \mathbf{z}| \mathbf{H}, n).
\end{equation}
\end{cor}

\corref{cor2.1.1} indicates that PJ can offer MLD performance in LQS-MIMO systems at high SNRs.
Now, we consider to derive the SER of PJ in MIMO systems.

\begin{cor} \label{cor2.1.2}
Suppose that the error occurs independently at each data stream for both PJ and $\bar{\mathbf{z}}$, denote $\mathcal{E}_{\textsc{pj}}$ as the SER of PJ, it can be expressed as follows
\begin{equation} \label{eqn27}
	\mathcal{E}_{\textsc{pj}} \leq \min \{\mathcal{E}_{\textsc{in}},\Psi\},
\end{equation}
where $\mathcal{E}_{\textsc{in}}$ denotes the SER of initialization detector as follows
\begin{equation} \label{eqn28}
	\mathcal{E}_{\textsc{in}} = \dfrac{1}{N} \sum_{k = 1}^{N}\mathcal{E}_{k}^{(\textsc{in})},
\end{equation} 
and $\Psi$ denotes the union bound of $\mathcal{E}_{\textsc{pj}}$ as follows
\begin{equation} \label{eqn29}
	\Psi = \mathcal{E}_{\textsc{mfb}} \prod_{k=1}^{N} \left(1 - \mathcal{E}_{k}^{(\textsc{in})}\right) + \dfrac{1}{N}\sum_{n=1}^{N}\sum_{k=1}^{N} \mathcal{E}_{n|k}^{(\textsc{pj})} \mathcal{E}_{k}^{(\textsc{in})},
\end{equation}
where $\mathcal{E}_{n|k}^{(\textsc{pj})}$ denotes the error probability of user $n$ conditioned on $\bar{e}_{k} \neq 0$;
$\mathcal{E}_{k}^{(\textsc{in})}$ denotes the error probability of $\bar{z}_{k}$.
Similar to \eqref{eqn20}, $\mathcal{E}_{n|k}^{(\textsc{pj})}$ and $\mathcal{E}_{k}^{(\textsc{in})}$ are approximated as follows
\begin{IEEEeqnarray}{ll}
\mathcal{E}_{n|k}^{(\textsc{pj})} &\approx \mathcal{K} \mathscr{P}_{\textsc{pj}}\left(\mathbf{x} \rightarrow \mathbf{z} | \mathbf{H}, n, k \right); \label{eqn30}\\
\mathcal{E}_{k}^{(\textsc{in})} &\approx \mathcal{K}\mathscr{P}_{\textsc{in}}\left(\mathbf{x} \rightarrow \bar{\mathbf{z}} | \mathbf{H}, k \right), \label{eqn31}
\end{IEEEeqnarray}
where $\mathscr{P}_{\textsc{in}}\left(\mathbf{x} \rightarrow \bar{\mathbf{z}} | \mathbf{H}, k \right)$ represents the PEP of the initialization detector that makes a single error at $\bar{z}_k$.
For instance, if the initialization detector is ZF, the PEP is given by \cite{Minango2018}
\begin{equation} \label{eqn32}
	\mathscr{P}_{\textsc{in}}\left(\mathbf{x} \rightarrow \bar{\mathbf{z}} | \mathbf{H}, k \right) = \mathcal{Q}\left( \sqrt{\dfrac{d^{2}_{\mathrm{min}}}{2 \sigma_{v}^{2} [\mathbf{A}^{-1}]_{k,k}}} \right),
\end{equation}
where $[\cdot]_{k,k}$ is the $k^{th}$ diagonal element of the input matrix.
\end{cor}

Due to the approximations in \eqref{eqn30} and \eqref{eqn31},  the right term in \eqref{eqn27} is actually an approximation of $\mathcal{E}_{\textsc{pj}}$, not a strictly upper bound.
This will be demonstrated in the following section.

\section{Simulation Results} \label{sec5}
In this section, the objectives are: {\it 1)} to demonstrate the performance coincidence of MFB, MLD, AWGN and PJ in LQS-MIMO systems, and {\it 2)} to show that the SERs of MFB and PJ derived in \secref{sec4} are tight with the simulation results.
The objectives set the following two experiments.

\begin{figure*}[htb]
	\centering
	\subfloat{%
		\label{fig01a}
		\includegraphics[width=0.47\textwidth]{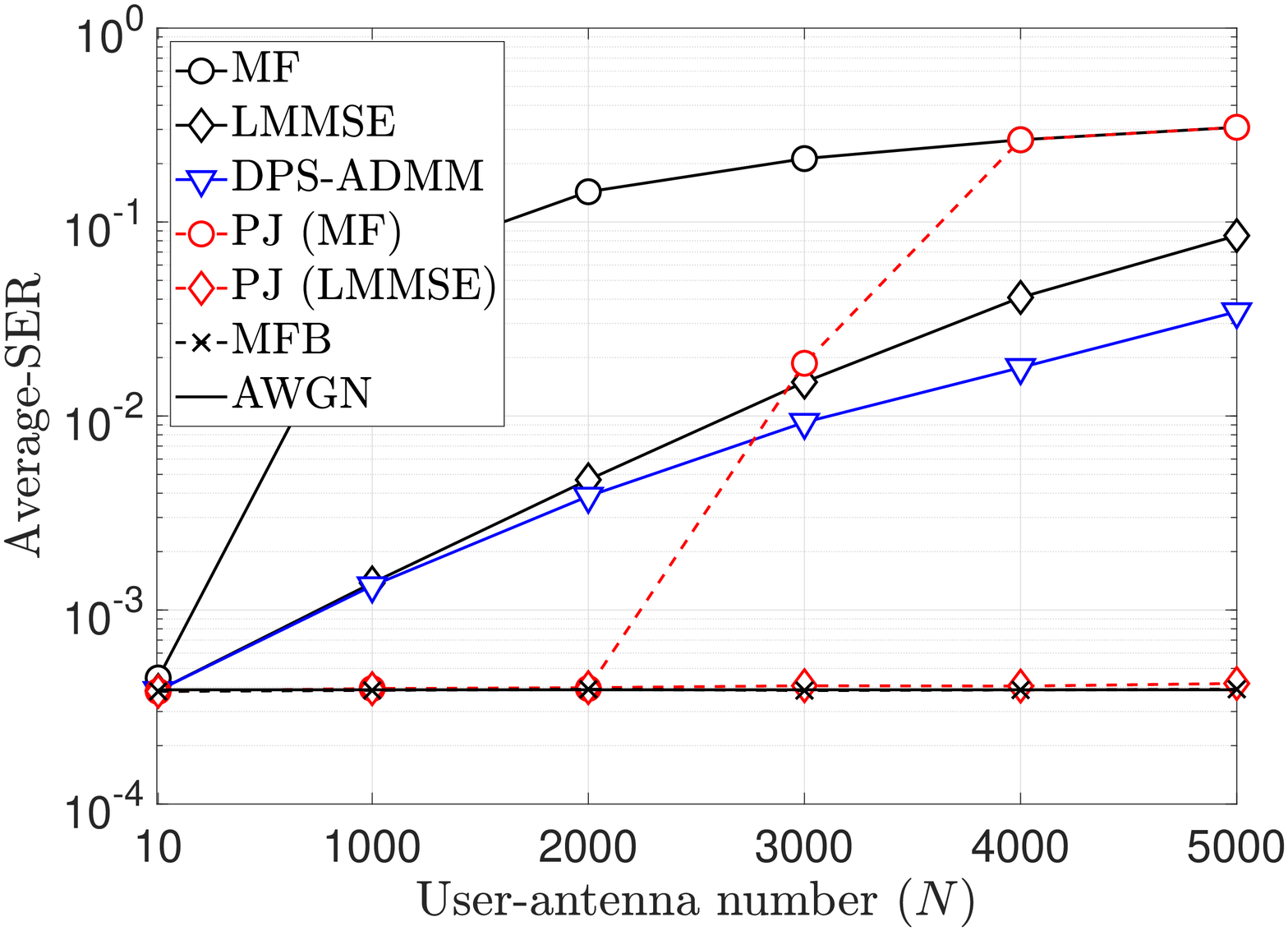}}
	\hfil
	\subfloat{%
		\label{fig01b}
		\includegraphics[width=0.47\textwidth]{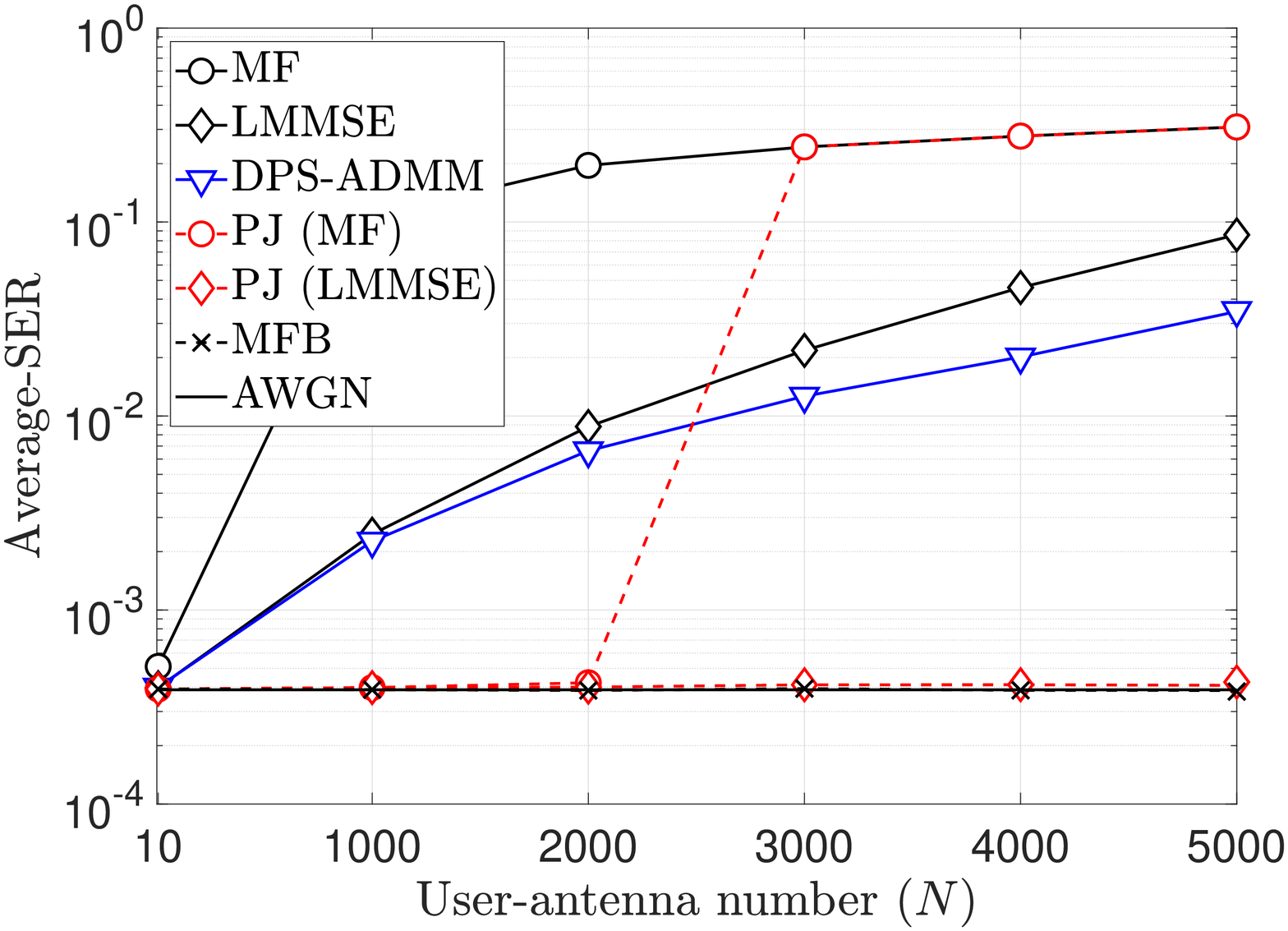}}  
	\caption{The average-SERs of PJ (MF) and PJ (LMMSE) in large MIMO systems as $N \rightarrow M$. $M = 5,000$;  $4$-QAM; Es/No = $11$ dB; $T = 5$. {\bf Left:} i.i.d. Rayleigh fading channels; {\bf Right:} i.n.d. Rayleigh fading channels.}
	\label{fig01}
	\vspace{-1em}
\end{figure*}
\begin{figure}[t]
	\centering
	\includegraphics[width=0.47\textwidth]{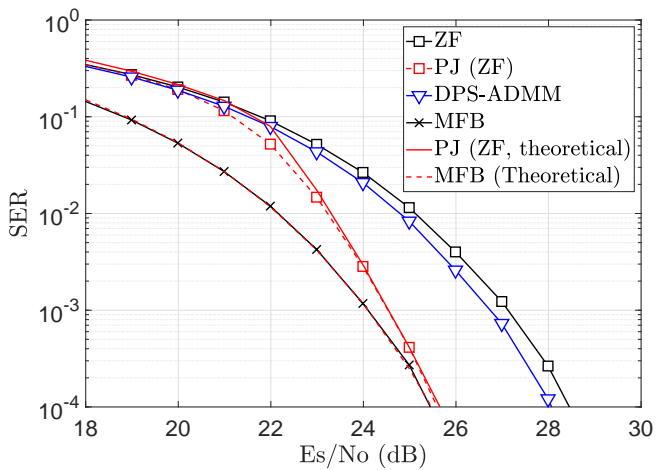}
	\caption{SERs of PJ and MFB in i.i.d. Rayleigh fading channel. $M = 128$; $N = 64$; $64$-QAM; $T = 3$.}
	\label{fig02}
	\vspace{-1em}
\end{figure}

{\it Experiment 1:}
In this experiment, we will show that PJ can offer the MLD (or MFB/AWGN) performance in LQS-MIMO systems.
The iteration number of PJ is set as $T=5$ and the modulation is $4$-QAM.
Furthermore, the initialization of PJ is set to be the decision of MF or LMMSE, where the latter can be obtained with square-order complexity.
A large \gls{ula} is deployed on the network-side, which has $M = 5,000$ service-antennas and the length of $214$ meters (on the $3.5$ GHz carrier frequency).
The users are located parallel to the ULA at a perpendicular distance of $50$ meters.
For non-stationary MIMO channels, the parameter of large-scale fading (i.e., $w_{m,n}$) is set to follow the statement in \cite{Amiri2018}.

\figref{fig01} shows that PJ (LMMSE) can achieve close-to-MFB/AWGN performance in (non-) stationary LQS-MIMO channels, even when $N/M = 1$.
This result implies that the detection performances of MLD and MFB are the same.
DPS-ADMM is proposed in \cite{Zhang2022}, which outperforms LMMSE, especially in symmetric MIMO systems.
PJ (MF) can offer close-to-optimum performance in MIMO systems, from low to medium load, i.e., $N \leq 2,000$.
This is inline with the discussion in \secref{sec03b}.
Moreover, when $N = 3000$, PJ (MF) has a performance degradation in the right sub-figure, due to the channel spatial non-stationarity.

{\it Experiment 2:}
In this experiment, we aims to show that the SERs derived in \corref{cor1.3.1} and \ref{cor2.1.2} can well fit the simulation results.
The modulation is set as $64$-QAM to show that PJ is scalable as $\mathcal{J}$ increases.
We consider an LQS-MIMO system with medium size and load, i.e., $M = 128$, $N = 64$.
The wireless channel follows i.i.d. Rayleigh distribution.
As shown in \figref{fig02}, PJ can still offer close-to-MFB detection performance, especially at high SNRs.
This is in line with our theoretical analysis in \secref{sec04b}.
Furthermore, it also shows that the simulation results match well with the mathematically derived SERs.
In i.n.d. Rayleigh channels, the performance of PJ (ZF) is similar to that in \figref{fig02}. 
Due to the page limit, such results are not present here.

\section{Conclusion} \label{sec6}
In this paper, PJ is proposed LQS-MIMO detection in both stationary and non-stationary channels.
PJ has square-order complexity and supports parallel computation, which is scalable for future LQS-MIMO systems.
It is theoretically proved that the detection performances of MLD, MFB and AWGN are asymptotically the same when $M \rightarrow \infty$.
Moreover, it is theoretically and empirically demonstrated that, PJ can achieve the MLD performance in LQS-MIMO systems, even when $N/M = 1$.
This means that an LQS-MIMO system using PJ can support massive user-antennas each with AWGN performance.
The SERs of MFB and PJ are mathematically derived and they are tight with the simulation results.

For practical large MIMO systems deployed  with an ELAA, the wireless channel can be more complicated.
For instance, there will be a mix of \gls{los} and non-LoS links with non-stationary shadowing effects in LQS-MIMO channels \cite{Liu2021}.
This kind of spatial non-stationarity makes the channel more ill-conditioned, and should be taken into account in future large MIMO transceiver design.

\section*{Acknowledgement}
This work is partially funded by the 5G Innovation Centre and 6G Innovation Centre.

\bibliographystyle{IEEEtran}
\bibliography{../IEEEabrv,../mMIMO} 

\end{document}